# Robust Transport of the Edge Modes along the Photonic Topological Interfaces of Different Configurations

J. Hajivandi and H. Kurt

Department of Electrical and Electronics Engineering

TOBB University of Economics and Technology, Ankara 06560, Turkey

**Abstract**

Two-dimensional photonic crystals made of six air holes on a core-shell dielectric material has been proposed to study the newly emerged photonic quantum spin Hall insulator. Specifically, radii modification of the air holes and core-shell without breaking time-reversal (TR) symmetry are supported by the $C_6$ point group symmetry upon a proposed scheme. It is shown that multiple topological transitions from an ordinary insulator with zero spin Chern number (Cs) to a topological insulator with a non-zero Cs can be achieved by modifying the geometry of the photonic structure. Studying the two counter-propagating helical edge modes which have the opposite group velocities are of individual importance for various optical purposes like scattering-free waveguides protected to various defects, disorders and strong light-matter interactions. We show that topological edge states demonstrate slow light characteristics. The findings emphasize the fact that exploring topological phase transition can be applied as a unique approach for realizing light transport, robust energy transportation and slow light in integrated photonic circuits and devices.

## 1. Introduction

Topology which is mainly the mathematics conception, related to the extremely strong preserved features is constant under the incessant modification of the bodily entities. To be more illustrative, one may consider the number of grips on an irregular surface. Then the genus as a robust topological feature may not be modified when the surface undergoes a continuous modification [1]. Since the first discovery of the remarkable behaviors of graphene in electronic structures, the topological insulators (TIs) as a new type of material which behaves as conductor at their edges but insulator in the bulk form, have been intensively studied [2]. Indeed, we can consider the quantum Hall effect in electronics providing topologically preserved one-way edge modes robust against the disorder, defects and cavity through the structure boundaries associated with protecting the topological band-gap. That feature is desired for various applications [3]. We can list different applications of TIs in numerous physical domains ranging from mechanics, acoustics, to microwaves [4].

Haldane and Raghu have extended the unusual Hall effect to the photonics field for the first time in the literature. Next, increasingly scientists concentrated on the photonic topological insulators (PTIs) by applying different types of photonic media especially photonic crystals (PCs) known as the semiconductor of electromagnetic waves. For example, different intriguing phenomena is realized in PTIs proposed by Wu and Hu in 2015 [5]. Indeed, if two PCs with the mutual band gap but different topological behavior are combined, then photonic edge modes which demonstrate topological features, emerge and propagate along the shared interface [6-14]. Moreover, the topological electromagnetic edge states may be induced in photonic medium demonstrating magneto-optic behavior and breaking the time-reversal (TR) symmetry. Indeed, coupling the magnetic and electric waves induces the bi-anisotropic effect of the metamaterial PCs [15]. Many researchers have explored various applications of this typical feature in numerous theoretical and experimental photonic purposes [16-31].

Besides, the triangular lattice PCs which follow the TR symmetry have been considered for more experimental perspectives because of their simple structure suitable for fabrication in comparison with the TR symmetry broken medium. As a result, the usages of Berry phases in photonic bands can be tailored by topological photonic modes [32]. Indeed, PCs preserving $C_6$ point group symmetry are regarded as unique types of PTIs supporting a four-folded degeneracy at the double Dirac cone appearing in the dispersion diagram. The topological edge modes would be produced by the pseudo-spin modes at the Dirac cone. Besides, the valley Hall effect is announced in optics. Reducing the $C_6$ point group symmetry in the graphene-like structures, so lifting the single Dirac degeneracy, lead to appearing two individual valley vortex modes with different chirality and emerging topological edge waves at the joint interface of trivial and non-trivial topological lattices. Besides, the valley Hall phononic crystals supporting acoustic waves, protect strong unidirectional transportation. Although, the robust propagation is observed at TIs including cavity, disorder and sharp corners, it is not obvious that when they act as valley Hall phononic crystals. Indeed, there are basic similarities and differences between TIs and valley Hall phononic crystals which can be implemented to design delay line devices and acoustic antennas [33-36].

In our recent works, the $C_6$ point group symmetry is reduced to $C_3$ one by implementing various modification on PC structure. Consequently, the double Dirac cone is lifted and two edge states at the band gap region are appeared which are robust and immune to scattering loss against numerous perturbations. It is illustrated that the topological Fano edge modes are emerged and preserved against the cavity and defects [37-40].

## 2. Design and Analysis of Photonic Structure

In this work, we theoretically propose a two-dimensional (2D) PCs for studying the topological phase transitions with TR invariance. Specifically, the structure with triangular-lattice 2D PC consists of core-shell dielectric materials connected by six air-holes. The schematic configuration of the studied structure is shown in Fig. 1.

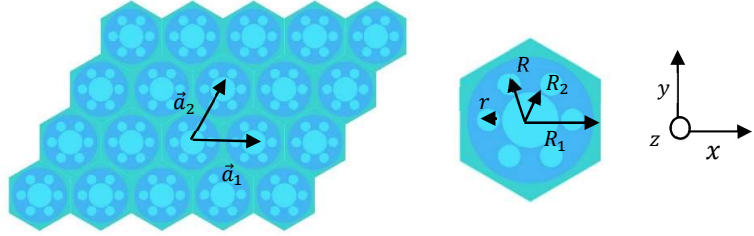

Fig 1. The schematic configuration of the triangular-lattice, 2D PC consists of core-shell dielectric materials (with the outer and inner radii $R_1$ and $R_2$ and the dielectrics $\varepsilon_1$ and $\varepsilon_2 = 1$ respectively) connected by micro-holes. $\vec{a}_1$ and $\vec{a}_2$ indicate the lattice vectors.

The lattice vectors are $\vec{a}_1 = (a, 0)$ and $\vec{a}_2 = \left(\frac{a}{2}, \frac{a\sqrt{3}}{2}\right)$ with the lattice constant $a$. Each hexagonal unit cell consists of core-shell dielectric materials (with outer and inner radii $R_1$ and $R_2$ and the dielectrics $\varepsilon_1$ and $\varepsilon_2 = 1$ respectively) surrounded by micro-hole with radius $r$ preserving $C_6$ symmetry. The distance between the adjacent air holes $R$, is also equal to the distance between an air hole and the center of a unit cell. This structure protects $Z_2$ PTIs. Based on the double-Dirac cone (four-folded degeneracy) at the central point of the Brillion zone, $\Gamma$, the quantum spin Hall effect can be realized. Indeed, double Dirac cone can be found by tuning the radii $R_1$ and $R_2$ as well as $R$ and $r$ which lead to inducing the topological properties and providing the trivial and nontrivial topological photonic states. Four-folded degeneracy at the $\Gamma$ point, composed of two two-folded dipolar states $(p_x, p_y)$ and quadrupole modes $d_{xy}/d_{x^2-y^2}$. The following pseudo-spin states can be defined based on the recognized dipolar and quadrupole states respectively,

$$p_\pm = \frac{p_x \pm i p_y}{\sqrt{2}}, \quad d_\pm = \frac{d_{x^2-y^2} \pm i d_{xy}}{\sqrt{2}}. \tag{1}$$

The TR operator, $T = UK = 1$ protects the pseudospin degeneracy. By employing the $k.p$ perturbation theory, the topological invariant due to the band inversion can be realized. The effective Hamiltonian can be deduced by pronouncing the dispersion at the $\Gamma$ point:

$$H(k) = \begin{pmatrix} \begin{pmatrix} M + Bk^2 & Ak_+ \\ A^* k_+ & -M - Bk^2 \end{pmatrix} & 0 \\ 0 & \begin{pmatrix} M + Bk^2 & Ak_- \\ A^* k_- & -M - Bk^2 \end{pmatrix} \end{pmatrix}, \tag{2}$$

with $k_\pm = k_x \pm i k_y$ & $M = \frac{\varepsilon_d - \varepsilon_p}{2}$ is negative (positive) in the topological (trivial) phases, ($\varepsilon_p, \varepsilon_d$ represent the dipolar and quadrupolar states at the $\Gamma$ point, respectively). Using the off-diagonal (diagonal) elements of the first (second) order perturbation term, $A$ ($B$) can be defined. The sign of the B is characteristically negative. So we can derive the Chern numbers as $[41-42]$:

$$C_\pm = \pm 1/2 [sign(M) + sign(B)]. \tag{3}$$

Therefore, by tuning the radii $R_1$ and $R_2$ as well as $R$ and $r$, the trivial and nontrivial topological photonic crystals (TPC and NTPC) can be realized. To illustrate that claim, we prepare Fig. 2 showing the photonic band diagrams and eigen-modes in two PCs with different topological states. We use plane-wave expansion method implemented in the *MIT Photonics Bands (MPB)* [43] for calculating the band structures of the 2D PCs for the TM polarization ($H_x, H_y, E_z \neq 0$) with the parameters: $\varepsilon_1 = 10.2$, $\varepsilon_2 = 1$, $R_1 = 0.5a$, $R_2 = 0.1a$, $R = 0.2a$ for two radii $r =$

$0.02a$ and $r = 0.08a$. Moreover, the dimensionless angular frequency $\frac{\omega a}{2\pi c}$ has been used where $c$ is the speed of the light in vacuum. As previously indicated there are doubly degenerated states for both the dipole, $p_\pm$ and the quadrupole, $d_\pm$ modes at the $\Gamma$ point. Actually by tuning the recognized radii, these two modes will change their place leading to the band inversion and two kinds of photonic insulators, trivial and non-trivial (topological) phases may appear. In the case of $r = 0.02a$, the quadrupole modes are above the dipole modes at the $\Gamma$ point which indicates normal band order (there is a small bandgap between orbitals $p$ and $d$ at the $\Gamma$ point which is not easily seen at the photonic band structure), whereas in the case of $r = 0.08a$, the quadrupole are below the dipole modes that is called parity-inversed band order indicating the topological characteristic of the PC.

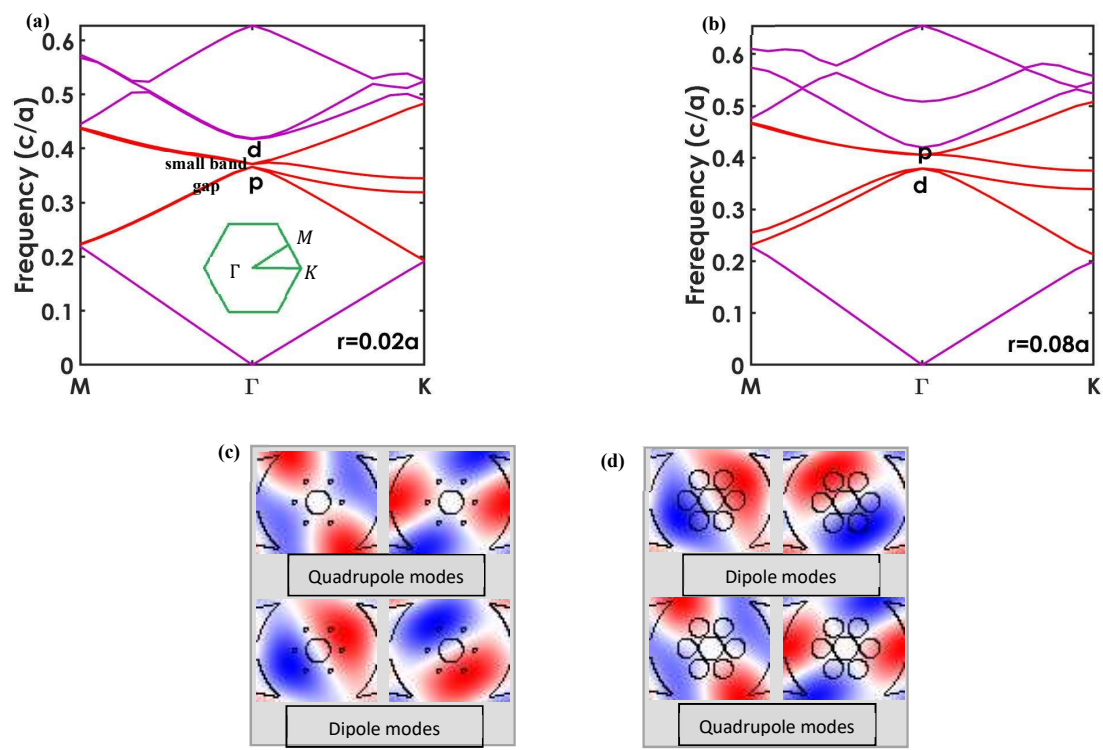

Fig. 2. Dispersion diagram and eigen-modes for the TM polarization in 2D PCs with $\varepsilon_1 = 10.2$, $\varepsilon_2 = 1.0$, $R_1 = 0.5a$, $R_2 = 0.1a$, $R = 0.2a$ for (a) $r = 0.02a$ (Inset: Brillouin zone of triangular lattice), (b) $r = 0.08a$. The magnetic field, $H_z$ profile of the dipole and quadrupole modes at the $\Gamma$ point. The quadrupole modes have higher energy than the dipole modes at the $\Gamma$ point for (c) $r = 0.02a$. The quadrupole modes have lower energy than the dipole modes at the $\Gamma$ point for (d) $r = 0.08a$.

In addition to band inversion, one may see the double-Dirac cone where the $p$ and $d$ states become degenerate at the $\Gamma$ point as shown in Fig. 3 for $r = 0.03a$. Actually the Dirac point as a kind of nodal point-where the different bands cross each other- is a critical point between trivial and non-trivial states. By applying parity/time modifications on the PC, we can reach to the trivial PC or topological band gap material, regarding to the distinct band gap perturbations [44].

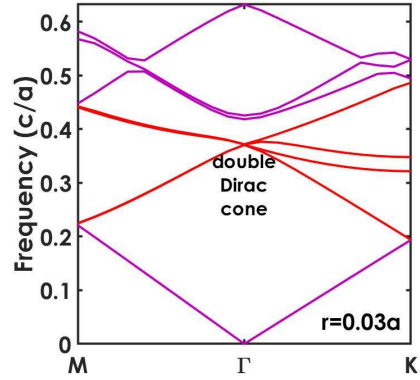

Fig. 3. Double Dirac cones appear at the Γ point when the radius $r = 0.03a$. The other parameters are the same as in Fig. 2.

It is possible to demonstrate that, more orbital states with lower energy ($s$) to higher energy ($f$) at the point Γ, appear at the band diagrams as presented in Fig. 4. These orbital modes are described at the Γ point for the $C_6$ symmetry [45]. In terms of the photonic bands, one can consider the Mie resonances resemble to the "atomic orbits", as it plays the same role as the atomic orbits in electronic energy bands [32,46,47].

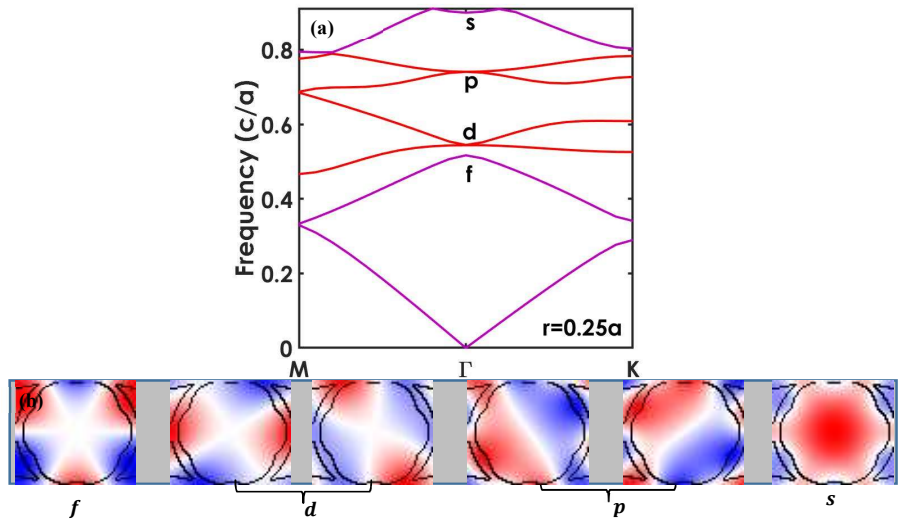

Fig. 4. (a) The band diagram of the 2D PCs with Eigen-modes $s$, $p$, $d$ and $f$ for the radius $r = 0.25a$. (b) The profiles of eigenmodes $s$, $p$, $d$ and $f$ for the radius $r = 0.25a$. The other parameters are the same as in Fig. 2.

By tuning the selected radii, the nodal points are appeared in the band diagram where the different bands cross each other. In Fig. 5, the triplet degeneracy between the double states $p$ and the singlet state $f$ is appeared which led to the quadratic dispersion relation in the band diagram. Similarity of the parity modes $s - d$ ($f - p$) leads to a quadratic dispersion characteristic touching points of the $s$ and $d$ ($f$ and $p$) bands while differences between the parity of modes $p$ and $s$ ($f$ and $d$) results in $p - s$ ($f - d$) degeneracy yielding linear dispersion [48, 49]. Such a nodal point including a flat band and a cone can be considered as a Dirac-like cone.

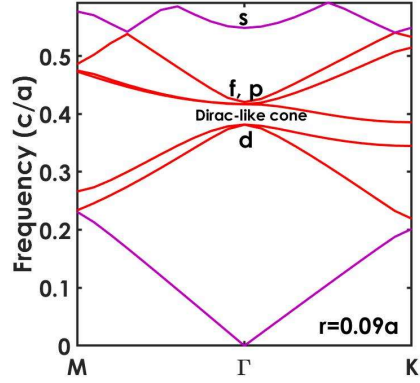

Fig. 5. Dirac-like cone at the central point of the Brillion zone Γ due to the degeneracy between orbitals $f$ and $p$, and other nodal points for the radius $r = 0.09a$. The other parameters are as the same as in Fig. 2.

In order to study the topological insulator behavior of PC, we choose the TPC with the appropriate parameters, $R_1 = 0.35a$, $R_2 = 0.1a$, $R = 0.2a$, $r = 0.13a$. In this case, $M = \frac{\varepsilon_d - \varepsilon_p}{2} > 0$ and $B < 0$ which results in $C_\pm = 0$. However, the PC with the parameters $R_1 = 0.45a$, $R_2 = 0.1a$, $R = 0.2a$ and $r = 0.15a$, shows non-trivial bandgap. It is because $(M,B) < 0$, leading to $C_\pm = \pm 1$.

### 3. Edge modes propagations

By considering a supercell of PTI which is infinite in one direction and composed of 17 and 6 topological and trivial unit cells in other directions, we study the helical edge states are appeared in the band structure diagram of the recognized system. It is clear that these two helical edge states are seen in the common photonic bandgap of two TPC & NTPC. These edge states (indicated with points A and B in the Fig. 6 (a)), have same frequency but opposite wave vectors. So they act as a Kramers pair and named as spin up (A) and spin down (B). To illustrate the unidirectional light propagation of the pseudo spin indicated edge states at the interface of the TPCs & NTPCs, we have designed an upper/lower system which occupied with the TPCs & NTPCs. The whole structure is surrounded by Perfectly Matching Layers (PMLs) in all sides to avoid reflecting electromagnetic waves into the simulation region. In order to study TM polarization, we consider an in-plane magnetic field of circular polarization for generating out of plane electric field of negative/positive angular momentum. To this aim we verify the helicity of the edge state by using two magnetic dipole source with 90-degree phase shift between them at the location indicated by the yellow star to propagate $E_z$ at the interface along the direction $+x$. We use Lumerical FDTD to perform the full wave simulation [50]. In order to check the robust transportation of the topological photonic edge state, we introduce specific defects including disorder and cavity into the system. As indicated in the Figs. 6(b)-(d), the edge states transport without noticeable back reflection, demonstrating robust unidirectional transport and topological protection along the straight interface, disorder and cavity respectively. The disorder perturbation is made by replacing two unit cells of TPC & NTPC which is shown in the yellow parallelogram and cavity is made by removing one-unit cell of the TPC shown in the yellow circle. The transmission of the edge state through the structures 6(b)-(d) are illustrated in the Fig. 6(e). We should note that in order to demonstrate the PTI light propagation investigated in time domain, we fixed lattice constant $a$=1000 nm and the corresponding frequency/wavelength region coinciding with the band diagram appears in the transmission window.

For studying the pseudospin-dependent light transportation, we employ a cross-shaped beam splitter. This cross-waveguide is separated to four subdivisions composed of TPCs & NTPCs. The labels from 1 to 4 indicate the input and output ports, respectively. Moving from port 1 to 2, as a straight waveguide, the NTPCs & TPCs are embedded in the right and left side exactly before the splitter region. After that, types of PCs are switching their positions immediately (NTPCs & TPCs on the left and right sides respectively) after the splitter region. As indicated in the Fig. 7(a), the light transport along the cross splitter depends on the spin of the source light. For example, the spin down edge state can support the right hand (clock-wise) transport propagating from port 1 to port 4. Indeed, the two edge states (pseudo-spin up and down) with opposite group velocities and rotations (right hand (clockwise)/left hand (counter-clockwise)) propagate in the opposite directions. This is a reminiscent of quantum spin Hall effect characterization. The simulated electric field transmission through the ports 1, 2, 3 and 4 are indicated in Fig. 7(b). Obviously, the light transmission from port 1 to port 2 is deeply suppressed. The splitter results for the transmission are overlapping with the unidirectional spin dependent light propagation in the PTIs.

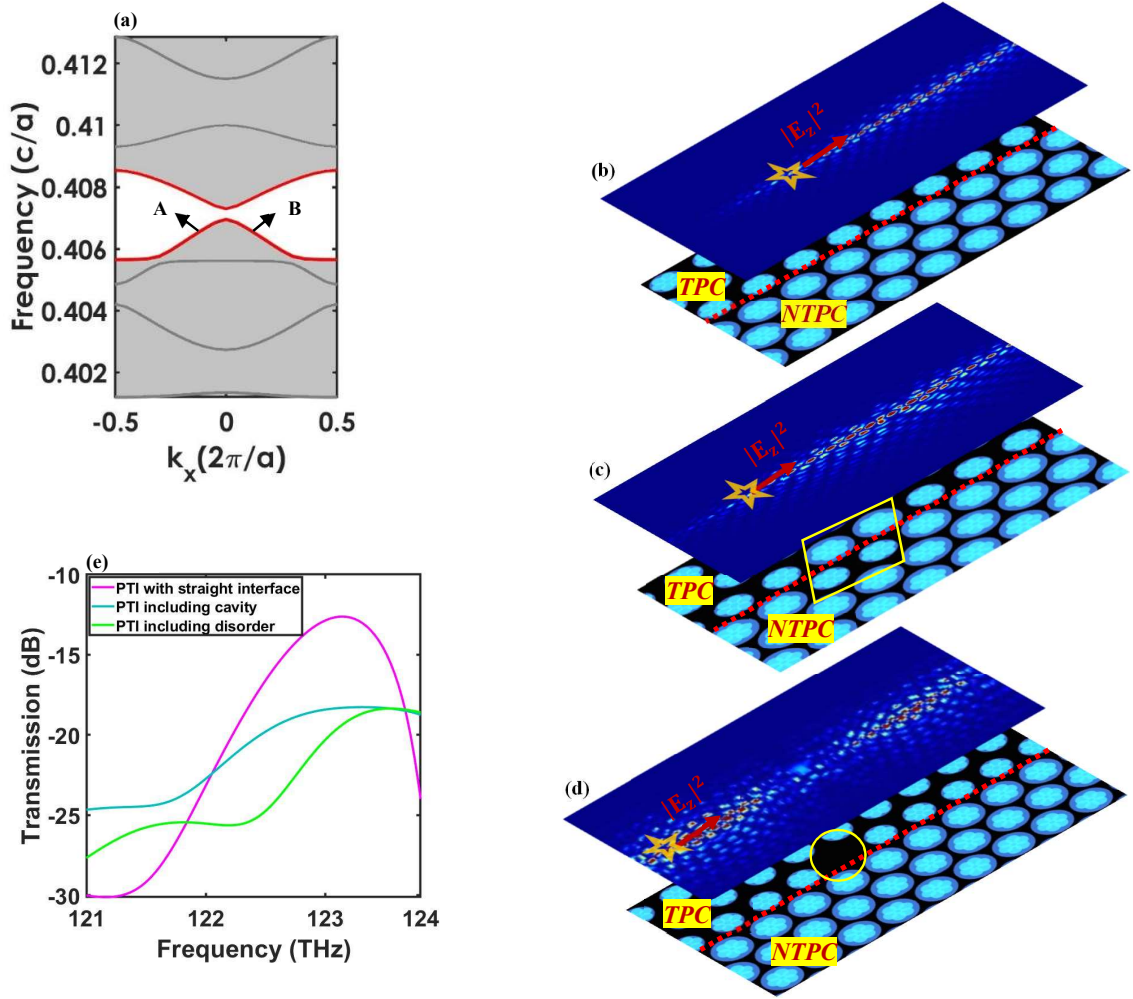

Fig. 6. (a) Dispersion diagram of a PTI with a supercell which is infinite in one direction and composed of 17 and 6 topological and trivial unit cells in other directions. The grey regions indicate the bulk photonic bands. Points A and B at the red ribbon-shaped represent the edge states. The edge mode back-scattering immune transport along the PTI (b) at straight interface, (c) including disorder and (d) against cavity, (e) transmission of the edge state for the structures given in (b)-(d).

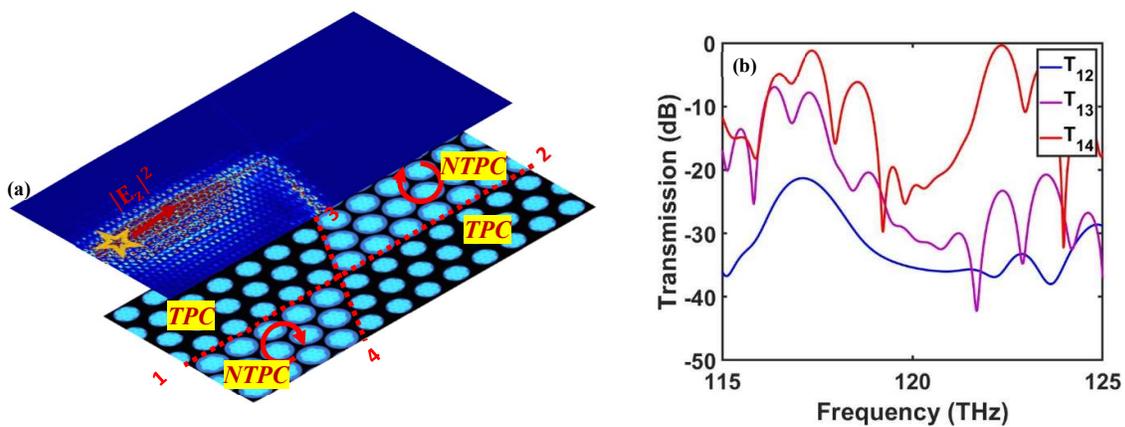

Fig. 7. (a) Clock-wise transportation of the edge states along the cross beam splitter including four ports 1-4, (b) transmission measured at ports 2, 3 and 4.

For studying the edge states transportation robustness against the defects, two kinds of disorders as well as two types of cavities are implemented into the cross waveguide splitter in Fig. 8 and Fig. 9, respectively. For making disorder type 1(2) or small (large) disorder, we replace one (two) unit-cells of TPC & NTPC along the straight interfaces. It can be found that the edge states pass through the defects keeping a high transmission through the port 1 to 4 rather than from the ports 1 to 2 or 1 to 3. Comparison between the transmission through the cross waveguide splitter not including the disorder and including the disorder 1 and 2, shows that it is almost not influenced by the presence of disorders as shown in Fig. 10 where different cases are collected at three output ports.

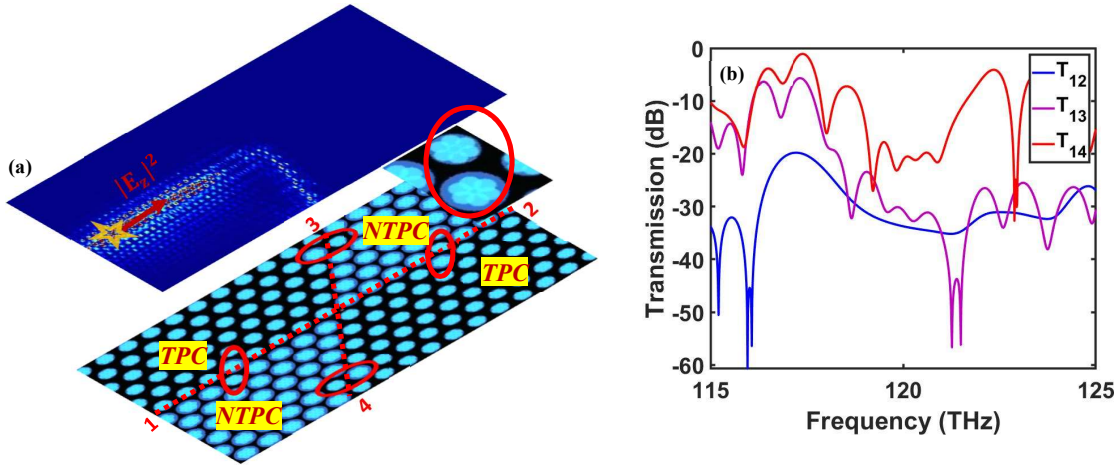

Fig. 8. (a) Clock-wise transportation of the edge states along the cross beam splitter including disorders (indicated by the red ovals) (b) transmission measured at ports 2, 3 and 4.

Cavity type 1(2) or small (big) can be created by removing one (two) unit-cells along the straight interfaces (Figs. 11 and 12). Again, high transmission of the edge states, from port 1 to 4 is detected. Besides, the transmission from port 1 to 2 is very low due to the clock-wise propagation of the edge state.

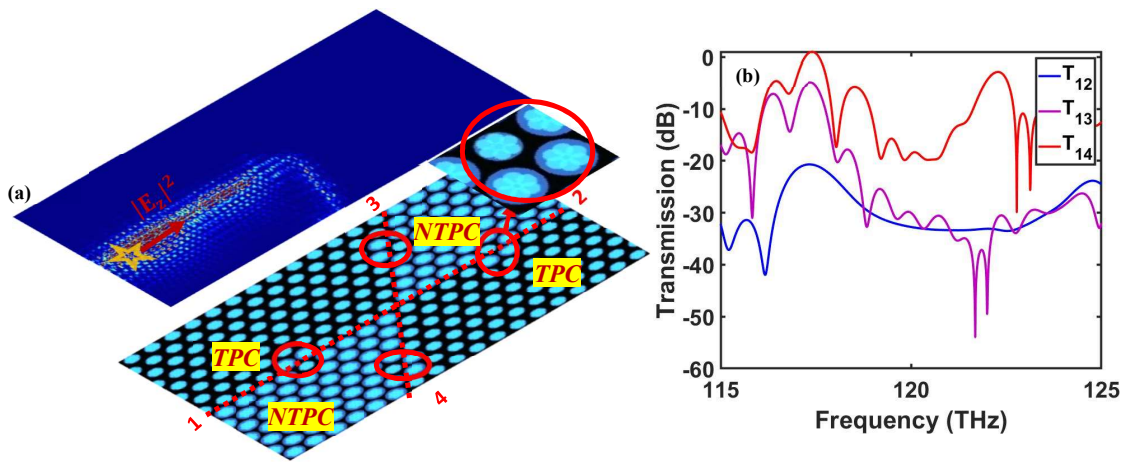

Fig. 9. (a) Clock-wise transportation of the edge states along the cross beam splitter including disorders (indicated by the red ovals) (b) transmission measured at ports 2, 3 and 4.

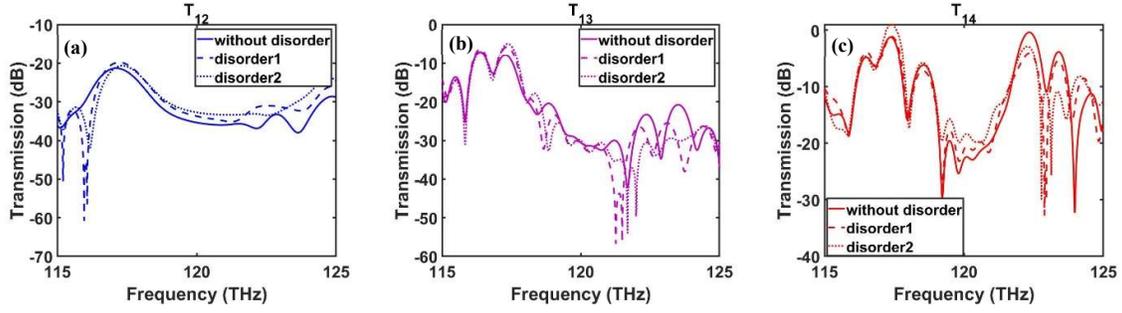

Fig. 10. Comparison between the transmission of the edge states for the PTI including straight interface, disorder 1 and 2.

Comparison between the transmission of the edge states for the small and large cavities are shows in Fig. 13. Hence, the robust light propagation is preserved through the topological cross waveguide. This novel topological PC, can be utilized for studying quantum characterization of classical modes. Besides, the established topological behaviors of light can be applied for the experimental applications.

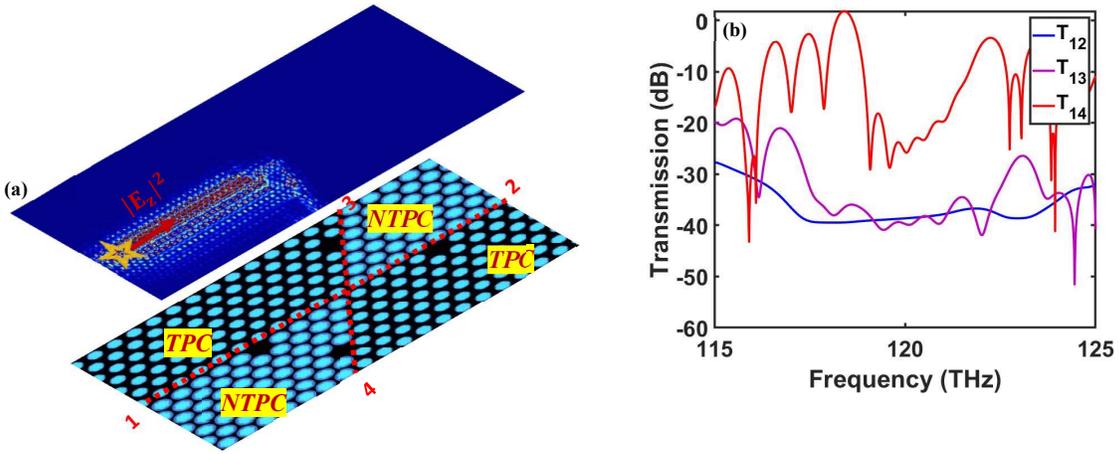

Fig. 11. (a) Clock-wise transportation of the edge states along the cross beam splitter including cavities (b) transmission measured at ports 2, 3 and 4.

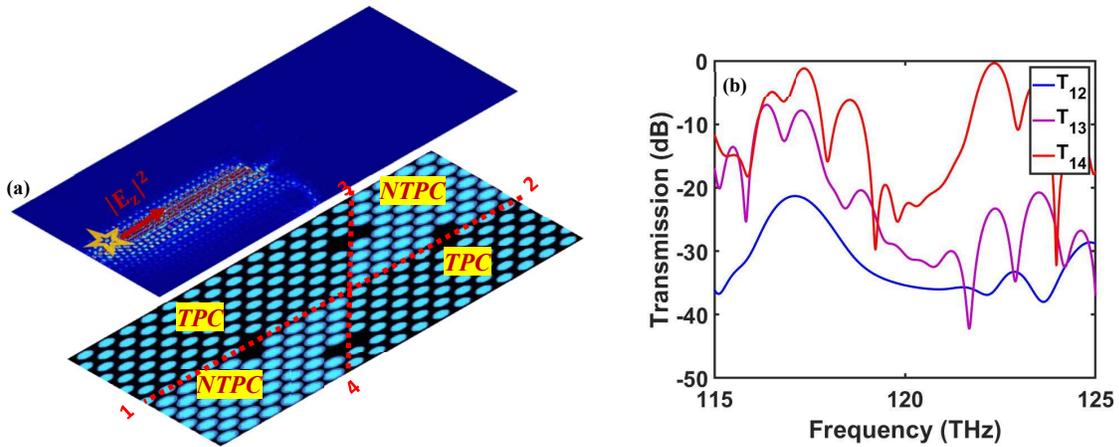

Fig. 12. (a) Clock-wise transportation of the edge states along the cross beam splitter including cavities (b) transmission measured at ports 2, 3 and 4.

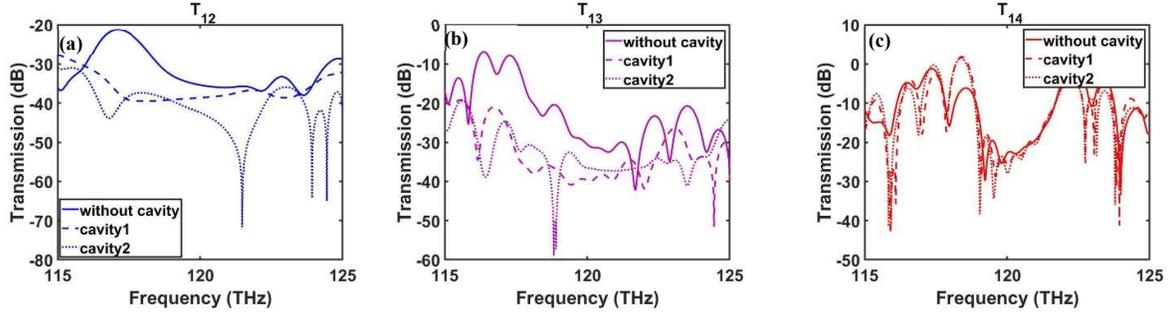

Fig. 13. Comparison between the transmission for the PTI including straight interface, cavity1 and 2.

We have shown the reliability of the edge modes against different kinds of defects or perturbations. This aspect is important to guide light under the presence of deformations. It is also crucial to underpin the group velocity of topological light guided between trivial and non-trivial phases. The group index is computed by FDTD analysis. A set of delay times at specific propagation positions is collected as presented in Fig. 14. The calculated group velocity of the edge states, $v_g$, and group index, $n_g$, are acquired as follows:

$$v_g = \frac{\Delta X}{\Delta t} = 0.365 \times 10^8 (m/s); \tag{4}$$

$$n_g = \frac{c}{v_g} = 8.219 ; \tag{5}$$

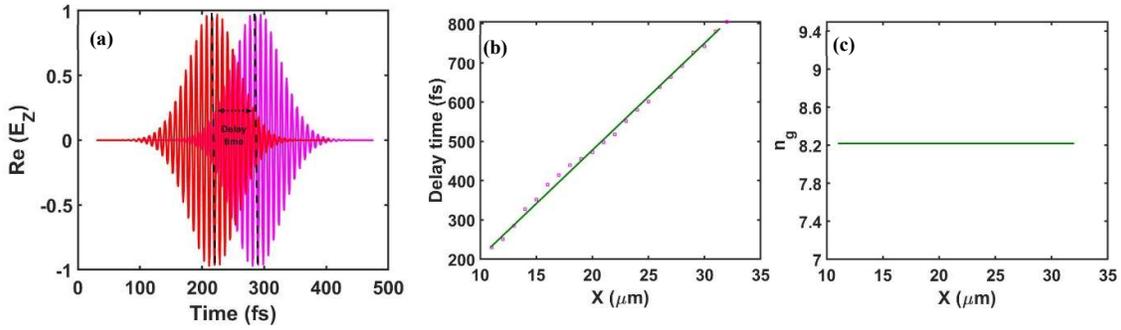

Fig. 14. (a) Time signal evolution of the edge state through the PTI interface, (b) delay time of the edge state through the PTI interface measured at different positions, (c) the equivalent group index.

The electric field variation in time domain through the straight PTI and PTI including a disorder are presented in Fig. 15. It can be found that, the edge state propagation through the disordered PTI has a delay time similar to the straight PTI. In addition, the edge mode propagates in the straight PTI with higher intensity in comparison with the disorder one. Besides, the transportation of edge wave packages become compressed in time domain in the case of disordered system.

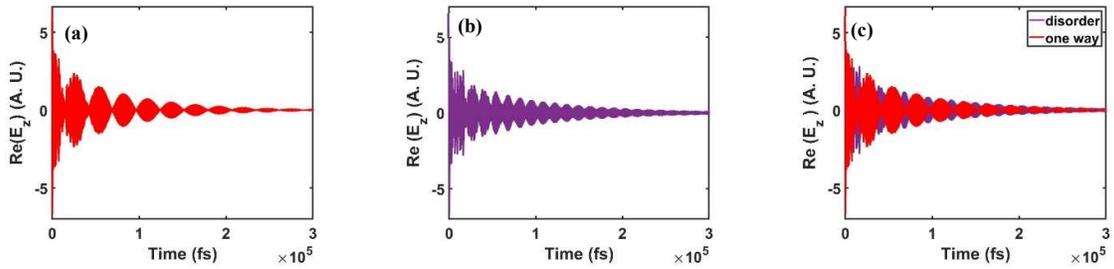

Fig. 15. (a) Time signal evolution of the edge state through the PTI interface for a long time duration, (b) time signal evolution of the edge state through the PTI including disorder for the same duration of time, (c) comparison between (a) and (b).

## 5. Conclusions

In conclusion, we have designed a new 2D PC with a hexagonal unit-cell preserving $C_6$ symmetry and consists of six air holes on a core-shell dielectric material. By tuning the radii of the holes and core-shell and without breaking the TR symmetry, topological properties of the PC can be performed in which the orbital $d$ becomes higher than orbital $p$ in the band structure diagram. Moreover, studying the photonic behavior of the edge states for the topological PC shows that although there is very little backscattering along the entire paths, it exhibits topological protection against distinct defects, cavities and disorders. By designing a cross waveguide composed of TPCs and NTPCs, we found that the light transportation along the different interfaces, depends on the spin of the edge state and is almost back-scattering immune and unaffected due to the disorder and cavities. It is shown and edge modes depict slow light behavior with group index value of approximately 8.2. As a result, the proposed PTI can be implemented for studying the quantum properties of the classical beams in simple and complex photonic structures.

## Acknowledgment

H. K. acknowledges partial support of the Turkish Academy of Sciences.